\documentclass[12pt,a4paper]{article}
\usepackage{amsmath}
\usepackage{graphicx}

\renewcommand{\baselinestretch}{2}

\begin{document}

\title{Optical transitions between Landau levels: AA-stacked bilayer graphene%
\\
}
\author{Yen-Hung Ho,$^{1,2}$ Jhao-Ying Wu,$^{1}$ Rong-Bin Chen,$^{3}$
Yu-Huang Chiu,$^{1,a)}$ \and and Ming-Fa Lin$^{1,b)}$ \\
{\small $^{1}$Department of Physics, National Cheng Kung University, Tainan
701, Taiwan}\\
{\small $^{2}$Department of Physics, National Sun Yat-Sen University,
Kaohsiung 804, Taiwan}\\
{\small $^{3}$Center of General Education, National Kaohsiung Marine
University, Kaohsiung 830, Taiwan}}
\date{\today}
\maketitle

\begin{abstract}
The low-frequency optical excitations of AA-stacked bilayer graphene are
investigated by the tight-binding model. Two groups of asymmetric LLs lead
to two kinds of absorption peaks resulting from only intragroup excitations.
Each absorption peak obeys a single selection rule similar to that of
monolayer graphene. The excitation channel of each peak is changed as the
field strength approaches a critical strength. This alteration of the
excitation channel is strongly related to the setting of the Fermi level.
The peculiar optical properties can be attributed to the characteristics of
the LL wave functions of the two LL groups. A detailed comparison of optical
properties between AA-stacked and AB-stacked bilayer graphenes is also
offered. The compared results demonstrate that the optical properties are
strongly dominated by the stacking symmetry. These optical properties could
be verified by optical measurements.

\vskip1.0 truecm $^{a)}$ Electronic mail: airegg.py90g@nctu.edu.tw \vskip0.0
truecm $^{b)}$ Electronic mail: mflin@mail.ncku.edu.tw \pagebreak

\pagebreak \renewcommand{\baselinestretch}{2}
\end{abstract}

\newpage \renewcommand{\baselinestretch}{2}

\section{Introduction}

Few-layer graphenes (FLGs) are very exotic nanomaterials owing to their
nanoscale interlayer distance, hexagonal symmetry, and stacking
configurations. FLGs have attracted numerous investigations on band
structures,$^{1-8}$ optical spectra,$^{1,9-17}$ electronic excitations,$%
^{5,18,19}$ and transport properties.$^{20,21}$ The presented properties
seem to make FLG-based materials excellent candidates for application in
electronic and photonic devices. Since the physical properties of FLGs are
strongly affected by the stacking configurations, FLGs with different
stacking configurations have attracted considerable experimental and
theoretical research. Monolayer (MG), AA-stacked bilayer (AABG) and
AB-stacked bilayer (BBG; bilayer Bernal) graphenes are three prototypical
FLGs. For the low-lying energy dispersions, MG exhibits isotropic linear
bands near the Fermi level ($E_{F}=0$); these bands become gradually
anisotropic parabolic bands as the energy exceeds the region of $\pm $0.5 eV.%
$^{2}$ In AABG, the linear bands in MG change into two pairs of linear
subbands with slightly different slopes.$^{5}$ These two pairs cross at $%
E_{F}=0$ and are asymmetric about $E_{F}=0$. For BBG, two pairs of parabolic
subbands exist that are asymmetric about $E_{F}=0$.$^{8,14}$ The conduction
and valence bands of the first pair slightly overlap about $E_{F}=0$.

In the presence of a uniform perpendicular magnetic field $\mathbf{B}=B_{0}%
\hat{z}$, the zero-field energy bands of FLGs become the dispersionless
Landau levels (LLs).$^{2-4,6-8}$ In MG, the low-lying LLs are characterized
by the special relation $E_{n}^{c,v}\propto \sqrt{n^{c,v}B_{0}}$, where $%
n^{c}$ ($n^{v}$) is the quantum number of the conduction (valence) LLs. This
special relation is broken as the energy exceeds the region of $\pm $0.5 eV
since the linear dispersions become gradually parabolic as the energy
exceeds this region.$^{2}$ For AABG and BBG, two groups of asymmetric LLs
exist.$^{8,14}$ These LLs are distributed away from a certain energy value
in each group. The LL energies of both bilayer graphenes do not exhibit a
simple relation similar to that of MG. These main features of FLGs would be
reflected in the magneto-optical absorption spectra. MG and BBG display
different optical properties, e.g., different field-dependent absorption
frequencies and distinct optical selection rules.$^{3,14}$

In this work, the magneto-optical absorption spectra of AABG are calculated
by gradient approximation$^{1,11,14,22}$ within the tight-binding model
(TB). Two groups of LLs are divided based on the characteristics of the LL
wave functions. The LL wave functions are clearly depicted and utilized to
explain the main features of the optical absorption spectra. The results
show that two kinds of absorption peaks exist in the absorption spectra.
Each peak obeys a single selection rule. The excitation channel associated
with each peak varies with a changing field strength. The optical properties
can be reasonably comprehended by the LL spectra and the characteristics of
the LL wave functions. A detailed comparison between AABG and BBG reveals
that they possess different magneto-optical absorption spectra, reflecting
the influences of stacking configurations on the electronic properties. In
other words, this method offers another way to distinguish AABG from BBG in
addition to the STM images. Moreover, it may also be used to discriminate
AABG from MG, which can be hardly done by STM.$^{23}$

\section{Landau level spectrum}

The geometric configuration of AABG is shown in Fig. 1(a). The primitive
unit cell consists of four sublattices, $A_{1}$, $B_{1}$, $A_{2}$, and $%
B_{2} $. The subscripts 1 and 2 are, respectively, the indices of the first
and second layer. Three atomic hopping integrals,$^{24}$ $\alpha _{0}$
(=2.569 eV), $\alpha _{1}$ (=0.361 eV), and $\alpha _{3}$ (=-0.032 eV), are
taken into account in this work. The first integral is the nearest-neighbor
hopping integral on the same layer and the second and third are the
interlayer interactions, as indicated in Fig. 1(a). $\mathbf{B}$ induces a
periodic Peierls phase related to the vector potential $\mathbf{A}{\mathbf{(%
\mathbf{r})}}=(0,B_{0}x,0)$. The Hamiltonian under a magnetic field is $H_{%
\mathbf{B}}={(\mathbf{P}}-{e\mathbf{A(\mathbf{r})}/c)^{2}}/{2m}+V(\mathbf{r}%
) $, where $m$, $\mathbf{P}$ ($=\hbar \mathbf{k}$), and $V(\mathbf{r})$ are
the electron mass, the crystal momentum, and the lattice potential,
respectively. Under the periodic condition, the primitive unit cell is
enlarged$^{2,7,8}$ and composed of four effective sublattices (denoted $%
A_{1} $, $B_{1}$, $A_{2}$, and $B_{2}$ for convenience, similar to the
symbols of the zero-field wave functions) including $2R_{B}$ $A_{1}$, $%
2R_{B} $ $B_{1}$, $2R_{B}$ $A_{2}$, and $2R_{B}$ $B_{2}$ atoms,
respectively. $R_{B} $ is defined by $R_{B}\equiv \Phi _{0}/({\frac{3\sqrt{3}%
}{2}B_{0}b^{2}})=\frac{79000\text{ T}}{{B_{0}}}$, where $\Phi _{0}$ is the
flux quantum and ${b}$ is the C-C bond length. $R_{B}$\ is inversely
proportional to $B_{0}$\ and related to the dimension of $H_{\mathbf{B}}$;$%
^{2,7,8}$ for example, $R_{B}$ is 1975 for $B_{0}=40$ T. That is to say,
each LL wave function is the linear combination of the four magnetic TB
functions associated with the four effective sublattices. The Hamiltonian
matrix elements in the presence of a magnetic field are
\begin{eqnarray}
\langle \mathbf{R}_{i,M}|H_{\mathbf{B}}|\mathbf{R}_{i^{\prime },M^{\prime
}}\rangle &=&\gamma _{s}(\mathbf{R}_{i,M}\text{, }\mathbf{R}_{i^{\prime
},M^{\prime }})\sum \frac{1}{\text{N}}\exp [i\mathbf{k}\cdot (\mathbf{R}%
_{i^{\prime },M^{\prime }}-\mathbf{R}_{i,M})  \notag \\
&&+i\frac{e}{\hbar }\int_{0}^{1}(\mathbf{R}_{i^{\prime },M^{\prime }}-%
\mathbf{R}_{i,M})\cdot \mathbf{A}[\mathbf{R}_{i^{\prime },M^{\prime
}}+\lambda (\mathbf{R}_{i^{\prime },M^{\prime }}-\mathbf{R}_{i,M})]d\lambda ]%
\text{,}
\end{eqnarray}%
where $\mathbf{R}_{i,M}$ is the position vector of $A_{i,M}$ or $B_{i,M}$. $%
\gamma _{s}$'s$(\mathbf{R}_{i,M}$, $\mathbf{R}_{i^{\prime },j^{\prime }})$
indicate the atomic interactions between the atoms at $\mathbf{R}_{i,M}$ and
$\mathbf{R}_{i^{\prime },M^{\prime }}$, i.e., they are $\alpha _{0}$, $%
\alpha _{1}$, and $\alpha _{3}$ in this work. The representation $A_{i,M}$ ($%
B_{i,M}$) indicates the $M$th ($M=1,2...2R_{B}$) $A$ ($B$) atom on the $i$th
($i=1,2$) layer.

The first Brillouin zone of AABG is shown in Fig. 1(b). For the low-energy
electronic structure of AABG, the linear subbands of MG change into two
pairs of linear subbands owing to the AA-stacking configuration and
interlayer interactions, as shown in Fig. 1(c). The analytical solution of
the energy dispersions of the first (second) pair can be described as $%
E^{c,v}=-\alpha _{1}\pm \frac{3}{2}\left( \ \alpha _{0}-\alpha _{3}\right)
bk+\frac{\alpha _{1}\alpha _{3}}{\alpha _{0}}$ ($E^{c,v}=\alpha _{1}\pm
\frac{3}{2}\left( \alpha _{0}+\alpha _{3}\right) bk+\frac{\alpha _{1}\alpha
_{3}}{\alpha _{0}}$),$^{5}$ where $E^{c}$ and $E^{v}$ are respectively
associated with the conduction and valence bands and are crossing at $%
-\alpha _{1}+\frac{\alpha _{1}\alpha _{3}}{\alpha _{0}}\simeq -0.366$ eV ($%
\alpha _{1}+\frac{\alpha _{1}\alpha _{3}}{\alpha _{0}}\simeq 0.357$ eV), as
indicated by the black (red) lines. $k$ is the wave vector measured from the
$\mathbf{K}$ point. The conduction and valence bands of each pair are
symmetric about the crossing energy. Furthermore, the two pairs of subbands
intersect at $E_{F}=0$ and their slopes are slightly different. This causes
the occupied and unoccupied states to be asymmetric about $E_{F}=0$. For the
wave functions of zero-field subbands (not shown), the first and second
pairs show the special relations, $A_{1}/A_{2}$ $=B_{1}/B_{2}=-1$ and $%
A_{1}/A_{2}$ $=B_{1}/B_{2}=1$, respectively. These relationships mainly
result from the nearest-neighbor interlayer interaction $\alpha _{1}$, which
leads to the separation of the first pair from the second pair.\ These main
features of the zero-field subbands would be reflected in a LL spectrum.

The magnetic field quantizes the two pairs of zero-field linear bands into
fourfold degenerate LLs, as shown in Fig. 2(a) for 40 T. Based on the
characteristics of the wave functions (Fig. 2(b)) discussed below, these LLs
can be further divided into two groups of LLs, $^{1st}$LLs and $^{2nd}$LLs.
The first (second) group corresponds to the first (second) pair of
zero-field subbands and is distributed away from the onset energy -0.366 eV
(0.357 eV), as shown in Fig. 2(a) by the black (red) lines. The onset energy
of the first (second) group is located at the crossing energy of the first
(second) pair of zero-field subbands. The LLs of each group are symmetric
about the onset energy of this group. However, the occupied and unoccupied
LLs are asymmetric about $E_{F}=0$. The LL energies of the two groups can be
written as
\begin{subequations}
\begin{eqnarray}
E_{n_{1}^{c,v}} &\simeq &-\alpha _{1}+\frac{\alpha _{1}\alpha _{3}}{\alpha
_{0}}\pm \frac{3}{2}\left( \ \alpha _{0}-\alpha _{3}\right) b\sqrt{%
2eB_{0}n_{1}^{c,v,eff}/\hbar } \\
E_{n_{2}^{c,v}} &\simeq &\alpha _{1}+\frac{\alpha _{1}\alpha _{3}}{\alpha
_{0}}\pm \frac{3}{2}\left( \ \alpha _{0}+\alpha _{3}\right) b\sqrt{%
2eB_{0}n_{2}^{c,v,eff}/\hbar }\text{,}
\end{eqnarray}%
where $n_{1}^{c,v,eff}$ and $n_{2}^{c,v,eff}$ are effective quantum numbers
of the LLs in the first and second groups and defined below. The low-lying
LL energies in each group linearly depend on $\sqrt{B}$ similar to the
relationship of MG.$^{2,3}$

The four degenerate wave functions of a LL are similar and thus only one of
them is discussed here. Each LL wave function of AABG is linearly combined
by four TB functions associated with the four effective sublattices. These
TB functions display oscillatory modes and localizaed features. The four TB
functions localized at a certain position are chosen for discussions (Fig.
2(b)).$^{8,14}$ Through appropriate fitting, the wave functions of the $n$th
$^{1st}$LL ($^{2nd}$LL) can be expressed as $A_{1}=-A_{2}\propto \varphi
_{n-1}\left( x\right) $ and $B_{1}=-B_{2}\propto \varphi _{n-2}\left(
x\right) $ ($A_{1}=A_{2}\propto \varphi _{n-1}\left( x\right) $ and $%
B_{1}=B_{2}\propto \varphi _{n-2}\left( x\right) $). $\varphi _{n}\left(
x\right) $ is the product of the $n$th-order Hermite polynomial and Gaussian
function,$^{3,8,11,14}$ where $n$ is the number of zero points of $\varphi
_{n}\left( x\right) $ and chosen to define the quantum number of a LL.$%
^{ref} $ For convenience, the zero-point numbers of the $A$ atoms are chosen
as effective quantum numbers of the LLs. $n_{1}^{eff}$'s and $n_{2}^{eff}$'s
are, respectively, the effective quantum numbers of the LLs in the first and
second groups, as shown in Fig. 2(a). Thus, the effective quantum numbers of
the $n$th conduction (valence) LLs in both groups are $n-1$, which are
similar to those of the LL wave functions of MG.$^{3}$ Furthermore, the LL
wave functions in Fig. 2(b) show a special relationship similar to those of
the zero-field wave functions, i.e., $A_{1}/A_{2}$ $=$ $B_{1}/B_{2}=-1$ and $%
A_{1}/A_{2}$ = $B_{1}/B_{2}$ $=1$ for the first and second groups,
respectively. These relations mainly originate from the specific stacking
configuration and should strongly affect the optical properties, e.g., the
optical-absorption peak structure and the optical selection rules. The LL
wave functions in both groups exhibit features similar to those of MG, i.e.,
they display similar oscillation modes, localization features, and a
combination of the $A$ and $B$ atoms in each layer. Obviously, the
characteristics of the LLs reflect those of the zero-field subbands, i.e.,
the existence of the two LL groups, the specific onset energies of the two
groups, the symmetric structure about the onset energy of each group, the
asymmetry of the occupied and unoccupied LLs about $E_{F}=0$, and the
specific relations of the LL wave functions of the two groups.

\section{Magneto-optical properties}

The main features of the LL spectra would be reflected in optical
excitations. At zero temperature, there exist only excitations from the
occupied to the unoccupied states. Based on Fermi's golden rule, the optical
absorption function is given by
\end{subequations}
\begin{eqnarray}
A(\omega ) &\propto &\sum_{h,h^{\prime }}\int_{1stBZ}{\frac{d\mathbf{k}}{%
(2\pi )^{2}}}|\langle \psi ^{h^{\prime }}(n_{1,2}^{h^{\prime },eff},\mathbf{k%
})|{\frac{\widehat{\mathbf{E}}\cdot \mathbf{P}}{m_{e}}}|\psi
^{h}(n_{1,2}^{h,eff},\mathbf{k})\rangle |^{2}  \notag \\
&&\times Im\{\frac{f\left[ E^{h^{\prime }}\left( \mathbf{k}%
,n_{1,2}^{h^{\prime },eff}\right) \right] -f\left[ E^{h}\left( \mathbf{k}%
,n_{1,2}^{h,eff}\right) \right] }{E^{h^{\prime }}\left( \mathbf{k}%
,n_{1,2}^{h^{\prime },eff}\right) -E^{h}\left( \mathbf{k},n_{1,2}^{h,eff}%
\right) -\omega -i\Gamma }\}\text{.}
\end{eqnarray}%
$\widehat{\mathbf{E}}$ is the unit vector of an electric polarization and $%
\widehat{\mathbf{E}}\parallel \widehat{x}$ is taken into account in this
work. $\Gamma $ (=1 meV) is a broadening parameter and often affected by
temperature and defect effects. $h$ and $h^{\prime }$ represent the occupied
and unoccupied states, respectively. $\langle \psi ^{h^{\prime
}}(n_{1,2}^{h^{\prime }},\mathbf{k})|\widehat{\mathbf{E}}\cdot \mathbf{P/}%
m_{e}|\psi ^{h}(n_{1,2}^{h},\mathbf{k})\rangle $ is the velocity matrix
element (denoted $M^{h^{\prime }h}$) derived from the dipole transition and
calculated by gradient approximation.$^{11,14}$ Through detailed
calculations, $M^{h^{\prime }h}$ is expressed as
\begin{equation}
\sum_{i,j=1,2}\sum_{M,M^{\prime }=1}^{2R_{B}}[A_{i}^{h^{\prime }\ast }\times
B_{j}^{h}]\nabla _{\mathbf{k}}\left\langle A_{i,M\mathbf{k}}\left\vert
H_{B}\right\vert B_{j,M^{\prime }\mathbf{k}}\right\rangle +h.c.
\end{equation}%
Eq. (4) corresponds to the two hopping integrals $\alpha _{0}$ (the terms
for $i=j$) and $\alpha _{3}$ (the terms for $i\neq j$), and the terms
associated with $\alpha _{0}$\ dominate the value of Eq. (4). For the sake
of convenience, the dominant term $A_{i}^{h^{\prime }\ast }\times B_{i}^{h}$
is represented by $M_{ii}^{h^{\prime }h}(\alpha _{0})$ in the following
discussions.

The low-frequency optical absorption spectra for $B_{0}=0$, $20$ T, and $40$
T, are shown in Fig. 3(a). The zero-field spectrum does not show any peak
structure and the absorption rate is zero as the energy is below $0.71$ eV.
The vanishing absorption rate corresponding to the intersubband excitations
is owing to the special relations of the wave functions in the two subbands.
The emergence frequencies of the intrasubband excitations associated with
the first and second pairs are nearly 0.73 eV ($\simeq 2\alpha _{1}-2\frac{%
\alpha _{1}\alpha _{3}}{\alpha _{0}}$; indicated by the black dot) and 0.71
eV ($\simeq 2\alpha _{1}+2\frac{\alpha _{1}\alpha _{3}}{\alpha _{0}}$;
indicated by the red dot), respectively. For $B_{0}=20$ T and $40$ T,
absorption peaks basically result from the excitations between two LLs in
the same group, i.e., only intragroup excitations exist in the absorption
spectrum. Each peak can clearly be identified. The excitation channels of
the first (second) kind of absorption peaks, $n_{1}^{c,v,eff}\rightarrow
n_{1}^{\prime c,v,eff}$ ($n_{2}^{c,v,eff}\rightarrow n_{2}^{\prime c,v,eff}$%
), are indicated by black (red) dots in Fig. 3(a). $n_{1}^{c,v,eff}%
\rightarrow n_{1}^{\prime c,v,eff}$ ($n_{2}^{c,v,eff}\rightarrow
n_{2}^{\prime c,v,eff}$) represents the excitation channel from the occupied
$^{1st}$LL with $n_{1}^{c,v,eff}$ ($^{2nd}$LL with $n_{2}^{c,v,eff}$) to the
unoccupied $^{1st}$LL with $n_{1}^{\prime c,v,eff}$ ($^{2nd}$LL with $%
n_{2}^{\prime c,v,eff}$). The $m$th absorption peak frequency of the first
(second) kind is denoted $\omega _{11}^{m}$ ($\omega _{22}^{m}$). For $%
B_{0}=20$ T ($40$ T), $\omega _{11}^{1}$ and $\omega _{22}^{1}$ originate
from $7^{c}\rightarrow 8^{c}$ and $8^{v}\rightarrow 7^{v}$ ($%
3^{c}\rightarrow 4^{c}$ and $4^{v}\rightarrow 3^{v}$) respectively, i.e.,
they originate in the excitations from the occupied conduction to the
unoccupied conduction LLs and from the occupied valence to the unoccupied
conduction LLs, respectively. $\omega _{11}^{1}$ and $\omega _{22}^{1}$ are
merged owing to their almost identical frequencies and strongly affected by
the setting of Fermi energy (the Fermi energy is set to $E_{F}=0$). Except
for $\omega _{11}^{1}$ and $\omega _{22}^{1}$, the other peaks, $\omega
_{11}^{m}$ and $\omega _{22}^{m}$ for $m\geq 2$, come from the excitations
between the occupied valence and unoccupied conduction LLs. For $B_{0}=20$ ($%
40$) T, the excitations $7^{v}\rightarrow 8^{c}$ and $8^{v}\rightarrow 7^{c}$
($3^{v}\rightarrow 4^{c}$ and $4^{v}\rightarrow 3^{c}$) result in $\omega
_{11}^{2}$ and $\omega _{22}^{2}$ respectively. For 20 T, the other
channels, $(m+5)^{v}\rightarrow (m+6)^{c}$ and $(m+6)^{v}\rightarrow
(m+5)^{c}$ possess the same frequency in the first (second) group which
result in $\omega _{11}^{m}$ ($\omega _{22}^{m}$). However, $\omega
_{11}^{m} $ ($\omega _{22}^{m}$) for 40 T originates from $%
(m+1)^{v}\rightarrow (m+2)^{c}$ and $(m+2)^{v}\rightarrow (m+1)^{c}$
possessing the same frequency in the first (second) group. Obviously, the
excitation channels of the $m$th peaks related to the different field
strengths may be different. $\omega _{11}^{m}$ and $\omega _{22}^{m}$ are
also slightly different. The former is higher than the latter, and they form
a pair-like structure. This reflects the asymmetric structure of the LLs.
The vanishing peak structure in the low-frequency region below $\omega
_{22}^{2}$, except for $\omega _{11}^{1}$ and $\omega _{22}^{1}$, can be
ascribed to the characteristics of the LL wave functions and is discussed
below. Simply said, the magneto-optical absorption spectra reflect the main
features of the zero-field spectrum, i.e., two kinds of absorption peaks
exist and only intragroup excitations are allowed.

The optical selection rules for $\omega _{11}$ and $\omega _{22}$ can be
represented by $\left\vert \triangle n_{11}\right\vert $ ($=\left\vert
n_{1}^{h^{\prime },eff}-n_{1}^{h,eff}\right\vert $)$=$ $\left\vert \triangle
n_{22}\right\vert $ ($=\left\vert n_{2}^{h^{\prime
},eff}-n_{2}^{h,eff}\right\vert $)$=1$. These rules are similar to those of
the LLs in MG. The selection rules can be comprehended by the
characteristics of the LL wave functions. In Eq. (3), $M_{ii}^{h^{\prime
}h}(\alpha _{0})$ dominates the excitations of the absorption peaks. $%
M_{ii}^{h^{\prime }h}(\alpha _{0})$ ($=$ $A_{i}^{h^{\prime }\ast }\times
B_{i}^{h}$) has non-zero values only when $A_{1}$ and $B_{1}$ ($A_{2}$ and $%
B_{2}$) own the same $\varphi _{n}(x)$ because of the orthogonality of $%
\varphi _{n}(x)$. Since $A_{1}$ ($A_{2}$) in the $n$th LL and $B_{1}$ ($B_{2}
$) in the $n+1$th LL for both groups own a same $\varphi _{n}(x)$, the
selection rules $\left\vert \triangle n_{11}\right\vert =$ $\left\vert
\triangle n_{22}\right\vert =1$ can be easily obtained. Furthermore, the
disappearance of the intergroup excitations can be ascribed to the fact that
the two products, $A_{1}\times B_{1}$ and $A_{2}\times B_{2}$, in the
intergroup excitations cancel each other out due to the special relationship
of the wave functions.

The field-dependent absorption frequencies associated with the first and
other $m$th peaks are shown in Figs. 3(b) and 3(c), respectively. In Fig.
3(b), at $B_{0}=60$ T, the excitation channel of $\omega _{11}^{1}$ ($\omega
_{22}^{1}$) comes from $2^{c}\rightarrow 3^{c}$ ($3^{v}\rightarrow 2^{v}$).
With decreasing field strength, the frequencies of $\omega _{11}^{1}$ and $%
\omega _{22}^{1}$ approach each other and merge at a sufficiently small
field strength. Furthermore, the excitation channels of the first peak are
altered as the field strength decreases to a critical field strength. For
example, the channel $2^{c}\rightarrow 3^{c}$ ($3^{v}\rightarrow 2^{v}$)
changes into $3^{c}\rightarrow 4^{c}$ ($4^{v}\rightarrow 3^{v}$) and $%
4^{c}\rightarrow 5^{c}$ ($5^{v}\rightarrow 4^{v}$) at $B_{0}=47$ T and $35$
T (indicated by two yellow lines), respectively. As $B_{0}$ is reduced, the
excitation channels become those associated with the LLs possessing larger
effective quantum numbers. The main reason for this is that the setting of
the Fermi level strongly affects which LL is considered the highest occupied
(lowest unoccupied) one. The discontinuity of field-dependent absorption
frequencies can be used in optical experiments to determine the excitation
channels. Moreover, the excitation frequency of each excitation channel in
its existent region is proportional to $B_{0}$, a behavior similar to the
excitation frequencies of MG. The other excitation channels shown in Fig.
3(c) show features similar to those in Fig. 3(b). The convergent frequencies
of $\omega _{11}$ and $\omega _{22}$ at the weak field strength are
approximately 0.73 eV and 0.71 eV and correspond to the two emergence
frequencies in the zero-field absorption spectrum, respectively. Since the
absorption peak intensities related to the LLs are strong, optical
measurements can reasonably determine the values of $\alpha _{1}$ and $%
\alpha _{3}$ through observing the convergent frequencies of absorption
peaks.

In addition to AABG, the magneto-optical properties of BBG are also
discussed in a previously published work.$^{14}$ AABG and BBG show similar
LL spectra, e.g., two groups of LLs and an asymmetric structure. However,
these two prototypical bilayer graphenes display totally different optical
properties. AABG exhibits two kinds of absorption peaks, and only intragroup
excitations that follow a single optical selection rule take place. BBG
produces four kinds of absorption peaks, and both intra- and inter-group
excitations that follow complex optical selection rules take place. These
optical properties can be comprehended by obtaining the characteristics of
the LL wave functions. The differences between the optical absorption
spectra of AABG and BBG imply that the optical properties can reflect the
influences of different stacking configurations. The dissimilarities between
these two graphenes are helpful for distinguishing AABG from BBG via optical
measurements. Furthermore, both the magneto-optical properties of AABG and
BBG reflect the main features of zero-field optical properties.

\section{Conclusions}

In summary, the TB calculations show that the LLs are asymmetric about $%
E_{F}=0$\ and can be divided into two groups based on the characteristics of
the wave functions. These two groups lead to two kinds of optical-absorption
peaks $\omega _{11}$'s and $\omega _{22}$'s associated with only intragroup
excitations of the first and second groups, respectively. The absorption
frequencies of $\omega _{11}^{m}$'s and $\omega _{22}^{m}$'s are slightly
different and form pair-like structures, which originate from the asymmetry
of LLs. The optical selection rules can be reasonably explained by the
characteristics of the LL wave functions. The selection rules of the two
kinds of peaks are $\triangle n_{11}=\triangle n_{22}=\pm 1$ similar to that
of MG. The similar selection rules of AABG and MG mainly originate from the
resembling characteristics of their LL wave functions. Furthermore, each $%
\omega _{11}^{m}$'s ($\omega _{22}^{m}$'s) corresponds to different
excitation channels within the different field strength region. The
field-dependent absorption frequencies for each excitation channel are
linearly dependent on $\sqrt{B_{0}}$, and resemble the absorption
frequencies of MG. The convergent absorption frequencies at the weak field
strength might be helpful and reliable in determining the interlayer atomic
interactions $\alpha _{1}$ and $\alpha _{3}$. The different magneto-optical
properties of AABG and BBG reflect the influences of different stacking
configurations. The differences between the two prototypical bilayer
graphenes can help experimental researchers discriminate AABG from BBG. The
above-mentioned magneto-optical properties could be confirmed by
magneto-absorption spectroscopy measurements.\newpage

\section{Appendix}

In the absence of external fields, the Hamiltonian is $H_{0}={\mathbf{P}^{2}}%
/{2m}+V(\mathbf{r})$, where $m$, $\mathbf{P}$ ($=\hbar \mathbf{k}$), and $V(%
\mathbf{r})$ are the electron mass, the crystal momentum, and the lattice
potential, respectively. In the presence of a uniform magnetic field $%
\mathbf{B}=B_{0}\widehat{z}$, the Hamiltonian is $H_{\mathbf{B}}={(\mathbf{P}%
}-{e\mathbf{A(\mathbf{r})}/c)^{2}}/{2m}+V(\mathbf{r})$, where ${\mathbf{A(%
\mathbf{r})}}=(0,B_{0}x,0)$ is the vector potential. A Peierls phase induced
by the magnetic field leads to a periodic condition and thus cause the
primitive uint cell to be enlarged, i.e., the primitive unit cell in the
absence of external fields is composed of four atoms, $A_{1}$, $B_{1}$, $%
A_{2}$ and $B_{2}$, while the enlarged primitive unit cell in the presence
of $\mathbf{B}$ comprises $2R_{B}$ $A_{1}$, $2R_{B}$ $B_{1}$, $2R_{B}$ $%
A_{2} $, and $2R_{B}$ $B_{2}$ atoms. $R_{B}\equiv \Phi _{0}/({\frac{3\sqrt{3}%
}{2}B_{0}b^{2}})=\frac{79000\text{ T}}{{B_{0}}}$ is associated with the
dimention of the Hamiltonian matrix. The Peierls is defined by
\begin{equation}
\Delta G=\int_{0}^{1}(\mathbf{R}_{i^{\prime },j^{\prime }}-\mathbf{R}%
_{i,j})\cdot \mathbf{A}[\mathbf{R}_{i^{\prime },j^{\prime }}+\lambda (%
\mathbf{R}_{i^{\prime },j^{\prime }}-\mathbf{R}_{i,j})]d\lambda \text{,}
\tag{A.1}
\end{equation}%
where $\mathbf{R}_{i,j}$ is the position vector of $A_{i,j}$ or $B_{i,j}$
for $i=1,2$ and $j=1,2,...2R_{B}$. In the sequence of the bases: $%
|A_{1}^{1}\rangle $, $|A_{1}^{2}\rangle $, $|B_{1}^{1}\rangle $, $%
|B_{1}^{2}\rangle $; $|A_{2}^{1}\rangle $, $|A_{2}^{2}\rangle $, $%
|B_{2}^{1}\rangle $, $|B_{2}^{2}\rangle $; ...; $|A_{2R_{B}}^{1}\rangle $, $%
|A_{2R_{B}}^{2}\rangle $, $|B_{2R_{B}}^{1}\rangle $, $|B_{2R_{B}}^{2}\rangle
$, the Hamiltonian matrix related to the enlarged unit cell is expressed by
\begin{equation}
H=\left(
\begin{array}{ccccccc}
H_{s,j=1} & H_{a,j=2} & 0 & 0 & \cdots & 0 & H_{d} \\
H_{a,j=2}^{\dag } & H_{s,j=2} & H_{a,j=3} & 0 & \cdots & 0 & 0 \\
0 & H_{a,j=3}^{\dag } & H_{s,j=3} & H_{a,j=4} & \ddots & 0 & 0 \\
0 & 0 & H_{a,j=4}^{\dag } & H_{s,j=4} & \ddots & 0 & 0 \\
\vdots & \vdots & \ddots & \ddots & \ddots & H_{a,j=2R_{B}-1} & 0 \\
0 & 0 & 0 & 0 & H_{a,j=2R_{B}-1}^{\dag } & H_{s,j=2R_{B}-1} & H_{a,j=2R_{B}}
\\
H_{d}^{\dag } & 0 & 0 & 0 & 0 & H_{a,j=2R_{B}}^{\dag } & H_{s,j=2R_{B}}%
\end{array}%
\right) \text{,}  \tag{A.2}
\end{equation}%
where
\begin{equation}
H_{s,j}=\left(
\begin{array}{cccc}
H_{A_{j}^{1}A_{j}^{1}} & H_{A_{j}^{1}A_{j}^{2}} & H_{A_{j}^{1}B_{j}^{1}} &
H_{A_{j}^{1}B_{j}^{2}} \\
H_{A_{j}^{2}A_{j}^{1}} & H_{A_{j}^{2}A_{j}^{2}} & H_{A_{j}^{2}B_{j}^{1}} &
H_{A_{j}^{2}B_{j}^{2}} \\
H_{B_{j}^{1}A_{j}^{1}} & H_{B_{j}^{1}A_{j}^{2}} & H_{B_{j}^{1}B_{j}^{1}} &
H_{B_{j}^{1}B_{j}^{2}} \\
H_{B_{j}^{2}A_{j}^{1}} & H_{B_{j}^{2}A_{j}^{2}} & H_{B_{j}^{2}B_{j}^{1}} &
H_{B_{j}^{2}B_{j}^{2}}%
\end{array}%
\right) =\left(
\begin{array}{cccc}
0 & \beta _{1} & \beta _{0}t_{1,j}^{\ast } & \beta _{3}t_{1,j}^{\ast } \\
\beta _{1} & 0 & \beta _{3}t_{1,j}^{\ast } & \beta _{0}t_{1,j}^{\ast } \\
\beta _{0}t_{1,j} & \beta _{3}t_{1,j} & 0 & \beta _{1} \\
\beta _{3}t_{1,j} & \beta _{0}t_{1,j} & \beta _{1} & 0%
\end{array}%
\right) \text{,}  \tag{A.3}
\end{equation}%
\begin{equation}
H_{a,j}=\left(
\begin{array}{cccc}
H_{A_{j-1}^{1}A_{j}^{1}} & H_{A_{j-1}^{1}A_{j}^{2}} &
H_{A_{j-1}^{1}B_{j}^{1}} & H_{A_{j-1}^{1}B_{j}^{2}} \\
H_{A_{j-1}^{2}A_{j}^{1}} & H_{A_{j-1}^{2}A_{j}^{2}} &
H_{A_{j-1}^{2}B_{j}^{1}} & H_{A_{j-1}^{2}B_{j}^{2}} \\
H_{B_{j-1}^{1}A_{j}^{1}} & H_{B_{j-1}^{1}A_{j}^{2}} &
H_{B_{j-1}^{1}B_{j}^{1}} & H_{B_{j-1}^{1}B_{j}^{2}} \\
H_{B_{j-1}^{2}A_{j}^{1}} & H_{B_{j-1}^{2}A_{j}^{2}} &
H_{B_{j-1}^{2}B_{j}^{1}} & H_{B_{j-1}^{2}B_{j}^{2}}%
\end{array}%
\right) =\left(
\begin{array}{cccc}
0 & 0 & 0 & 0 \\
0 & 0 & 0 & 0 \\
\beta _{0}q & \beta _{3}q & 0 & 0 \\
\beta _{3}q & \beta _{0}q & 0 & 0%
\end{array}%
\right) \text{,}  \tag{A.4}
\end{equation}%
and
\begin{equation}
H_{d}=\left(
\begin{array}{cccc}
H_{A_{1}^{1}A_{2R_{B}}^{1}} & H_{A_{1}^{1}A_{2R_{B}}^{2}} &
H_{A_{1}^{1}B_{2R_{B}}^{1}} & H_{A_{1}^{1}B_{2R_{B}}^{2}} \\
H_{A_{1}^{2}A_{2R_{B}}^{1}} & H_{A_{1}^{2}A_{2R_{B}}^{2}} &
H_{A_{1}^{2}B_{2R_{B}}^{1}} & H_{A_{1}^{2}B_{2R_{B}}^{2}} \\
H_{B_{1}^{1}A_{2R_{B}}^{1}} & H_{B_{1}^{1}A_{2R_{B}}^{2}} &
H_{B_{1}^{1}B_{2R_{B}}^{1}} & H_{B_{1}^{1}B_{2R_{B}}^{2}} \\
H_{B_{1}^{2}A_{2R_{B}}^{1}} & H_{B_{1}^{2}A_{2R_{B}}^{2}} &
H_{B_{1}^{2}B_{2R_{B}}^{1}} & H_{B_{1}^{2}B_{2R_{B}}^{2}}%
\end{array}%
\right) =\left(
\begin{array}{cccc}
0 & 0 & \beta _{0}q^{\ast } & \beta _{3}q^{\ast } \\
0 & 0 & \beta _{3}q^{\ast } & \beta _{0}q^{\ast } \\
0 & 0 & 0 & 0 \\
0 & 0 & 0 & 0%
\end{array}%
\right) \text{.}  \tag{A.5}
\end{equation}%
$t_{1,j}$ and $q$ are expressed as
\begin{subequations}
\begin{align}
t_{1,j}& =\exp \{i[-k_{x}{\frac{b}{2}}-k_{y}{\frac{\sqrt{3}b}{2}}+\pi {\frac{%
\phi }{\phi _{0}}}(j-1+{\frac{1}{6}})]\}  \notag \\
& +\exp \{i[-k_{x}{\frac{b}{2}}+k_{y}{\frac{\sqrt{3}b}{2}}-\pi {\frac{\phi }{%
\phi _{0}}}(j-1+{\frac{1}{6}})]\}\text{,}  \tag{A. 6(a)} \\
q& =\exp [ik_{x}b]\text{.}  \tag{A. 6(b)}
\end{align}

The optical absorption rate is dominated by the velocity matrix $%
M^{h^{\prime }h}$. Through the gradient approximation, the velocity matrix
is simplified as the product of three matrices, the initial state (occupied
state; $\psi ^{h}$), final state (unoccupied state; $\psi ^{h^{\prime }}$),
and the first-order differential of the Hamiltonian matrix element versus
the wave vector $\mathbf{k}$ ($\bigtriangledown _{\mathbf{k}}H_{ij}$). The
last term corresponds to the direction of electric polarization, i.e., it is
$\partial H_{ij}/\partial k_{x}$ ($\partial H_{ij}/\partial k_{y}$) for
polarization along the $x$-direction ($y$-direction). The elements in the
third matrix are non-zero only for $H_{ij}$ associated with the hopping
integrals, as are the velocity matrix elements. In other words, when the
velocity matrix element does not vanish, the initial and final states in the
product should be the two states in a hopping process corresponding to
non-vanishing hopping integrals. In this work, the polarization is along the
$x$-direction and $M^{h^{\prime }h}$ is expressed as
\end{subequations}
\begin{subequations}
\begin{multline}
M^{h^{\prime }h}=\left\langle \psi ^{h^{\prime }}\left\vert \partial
H/\partial k_{x}\right\vert \psi ^{h}\right\rangle   \notag \\
=\sum_{i,j=1,2}\left\langle \psi _{i}^{h^{\prime }}\left\vert
\sum_{M,M^{\prime }=1}^{2R_{B}}\partial /\partial k_{x}\left\langle \psi
_{i,M\mathbf{k}}^{h^{\prime }}\left\vert H_{B}\right\vert \psi _{j,M^{\prime
}\mathbf{k}}^{h}\right\rangle \right\vert \psi _{j}^{h}\right\rangle +h.c.%
\text{,}  \tag{A. 7}
\end{multline}%
where $\psi $ is the wave functions of the $A$ or $B$ atom. Eq. (A.7) is
associated with the three hopping integrals $\alpha _{0}$, $\alpha _{1}$ and
$\alpha _{3}$. The term related to $\alpha _{1}$ is vanishing since the
relative differential value is zero. $M^{h^{\prime }h}$\ can be simplified
as
\end{subequations}
\begin{subequations}
\begin{align}
& \sum_{i=1,2}\left\langle A_{i}^{h^{\prime }}\left\vert
\sum_{M=1}^{2R_{B}}\partial /\partial k_{x}\left\langle A_{i,M\mathbf{k}%
}\left\vert H_{B}\right\vert B_{i,M\mathbf{k}}\right\rangle \right\vert
B_{i}^{h}\right\rangle +h.c.  \notag \\
& +\sum_{i=1,2;\text{ }i\neq j}\left\langle A_{i}^{h^{\prime }}\left\vert
\sum_{M,M^{\prime }=1}^{2R_{B}}\partial H/\partial k_{x}\left\langle A_{i,M%
\mathbf{k}}\left\vert H_{B}\right\vert B_{j,M^{\prime }\mathbf{k}%
}\right\rangle \right\vert B_{j}^{h}\right\rangle \delta _{M,M^{\prime }\pm
1}+h.c.  \tag{A. 8 (a)} \\
& =M_{ii}^{h^{\prime }h}(\alpha _{0})+M_{ij}^{h^{\prime }h}(\alpha _{3})%
\text{.}  \tag{A. 8(b)}
\end{align}%
The first term, $M_{ii}^{h^{\prime }h}(\alpha _{0})$, is associated with $%
\alpha _{0}$ and the second term, $M_{ij}^{h^{\prime }h}(\alpha _{3})$, corresponds $\alpha _{3}$; $M_{ii}^{h^{\prime }h}(\alpha _{0})$ dominates
the value of the velocity matrix.

\newpage {\Large \textbf{References}} \newline

\begin{itemize}
\item[$^{1}$] C. P. Chang et al., Carbon \textbf{42}, 2975 (2004).

\item[$^{2}$] J. H. Ho et al., Physica E \textbf{40}, 1722 (2008).

\item[$^{3}$] Y. Zheng and T. Ando, Phys. Rev. B \textbf{65}, 245420 (2002).

\item[$^{4}$] E. McCann and V. I. Fal'ko, Phys. Rev. Lett. \textbf{96},
086805 (2006).

\item[$^{5}$] J. H. Ho et al., Phys. Lett. A \textbf{352}, 446 (2006).

\item[$^{6}$] E. V. Castro et al., Phys. Rev. Lett. \textbf{99}, 216802
(2007).

\item[$^{7}$] Y. H. Lai et al., Phys. Lett. A \textbf{372}, 292 (2008).

\item[$^{8}$] Y. H. Lai et al., Phys. Rev. B \textbf{77}, 085426 (2008).

\item[$^{9}$] M. L. Sadowski et al., Phys. Rev. Lett. \textbf{97}, 266405
(2006).

\item[$^{10}$] D. S. L. Abergel et al., Appl. Phys. Lett. \textbf{91},
063125 (2007).

\item[$^{11}$] Y. H. Chiu et al., Phys. Rev. B \textbf{78}, 245411 (2008).

\item[$^{12}$] C. L. Lu et al., Appl. Phys. Lett. \textbf{89}, 221910 (2006).

\item[$^{13}$] Z. Jiang et al., Phys. Rev. Lett. \textbf{98}, 197403 (2007).

\item[$^{14}$] Y. H. Ho et al., ACS Nano \textbf{4}, 1465 (2010).

\item[$^{15}$] E. A. Henriksen et al., Phys. Rev. Lett. \textbf{104}, 067404
(2010).

\item[$^{16}$] Y. Xu et al., Nanotechnology \textbf{21}, 065711 (2010).

\item[$^{17}$] Y. C. Lin et al., Appl. Phys. Lett. \textbf{96}, 133110
(2010).

\item[$^{18}$] J. H. Ho et al., Phys. Rev. B \textbf{74}, 085406 (2006).

\item[$^{19}$] X. F. Wang and T. Chakraborty, Phys. Rev. B \textbf{75},
041404 (2007).

\item[$^{20}$] Y. Zhang et al., Nature \textbf{438}, 201 (2005).

\item[$^{21}$] K. S. Novoselov et al., Science \textbf{315}, 1379 (2007).

\item[$^{22}$] M. F. Lin and Kenneth W.-K. Shung, Phys. Rev. B \textbf{50},
17744 (1994).

\item[$^{23}$] Z. Liu et al., Phys. Rev. Lett. \textbf{102}, 015501 (2009).

\item[$^{24}$] J.-C. Charlier and J.-P. Michenud, Phys. Rev. B \textbf{46},
4531 (1992).

\end{itemize}

\newpage {\Large \textbf{Figure Captions}}\newline

\begin{itemize}
\item[FIG. 1.] (a) The geometric structure, (b) the first Brillouin zone,
and (c) the low-lying subbands of AA-stacked graphene. $\alpha _{0}$ ($2.598$
eV) is the nearest-neighbor hopping integral and the two important
interlayer interactions are $\alpha _{1}$ ($0.361$ eV) and $\alpha _{3}$ ($%
-0.032$ eV).

\item[FIG. 2.] (a) Landau levels and (b) Landau level wave functions of
AA-stacked graphene at $B_{0}=40$ T. $n_{1}^{c,v}$'s and $n_{2}^{c,v}$'s are
the effective quantum numbers of the first and second group of Landau
levels, respectively.

\item[FIG. 3.] (a) The optical absorption spectra of AA-stacked graphene at $%
B_{0}=0$, $20$, and $40$ T. The field-dependent absorption frequencies
associated with (a) the first and (b) other $m$th absorption peaks. The
symbols $n\rightarrow n^{\prime }$ describe the excitation channels of the
absorption peaks.
\end{itemize}

\newpage
\begin{figure}[tbp]
\rotatebox{0}{\includegraphics[width=1\textwidth]{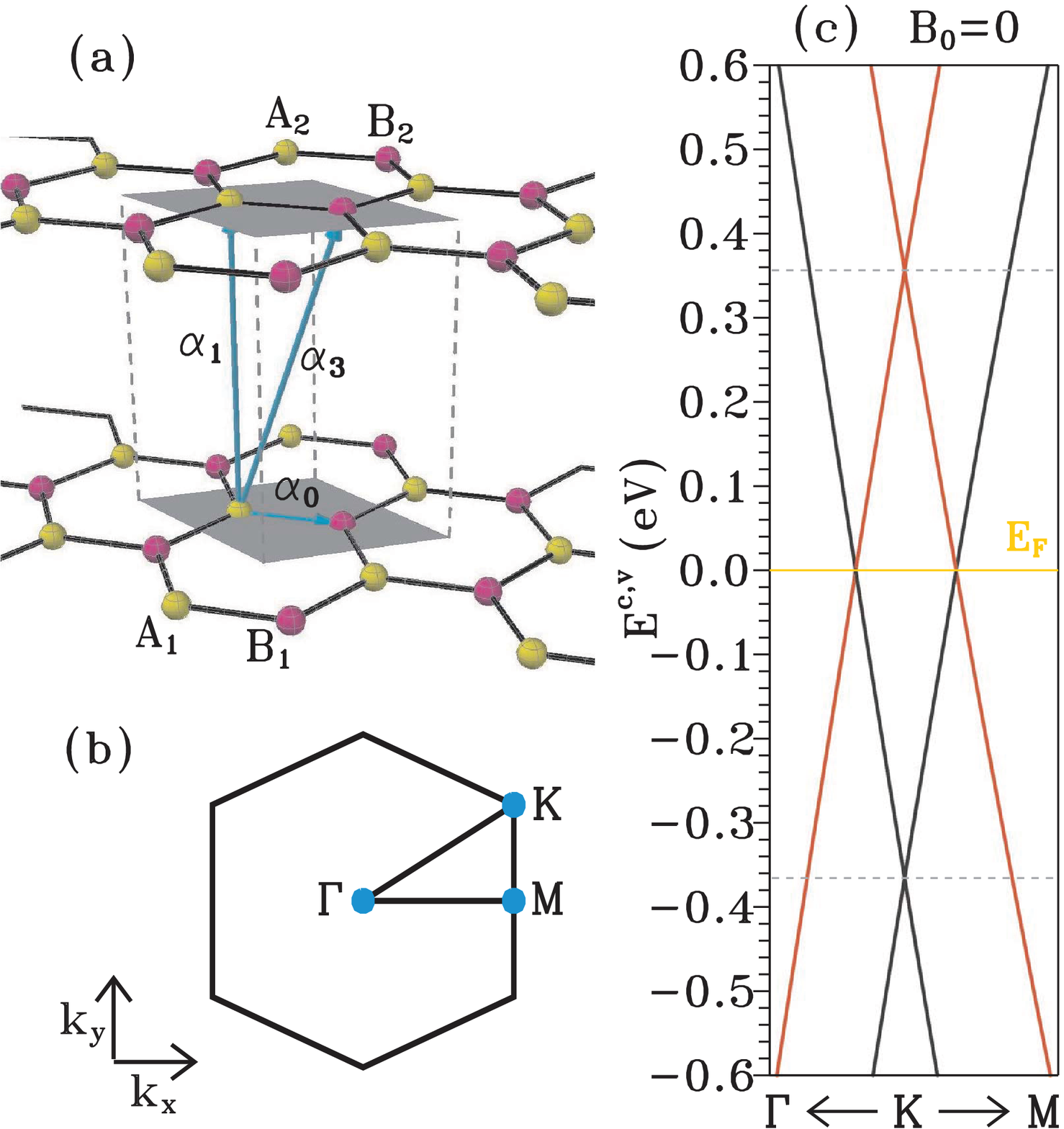}}
\end{figure}

\newpage
\begin{figure}[tbp]
\rotatebox{0}{\includegraphics[width=1\textwidth]{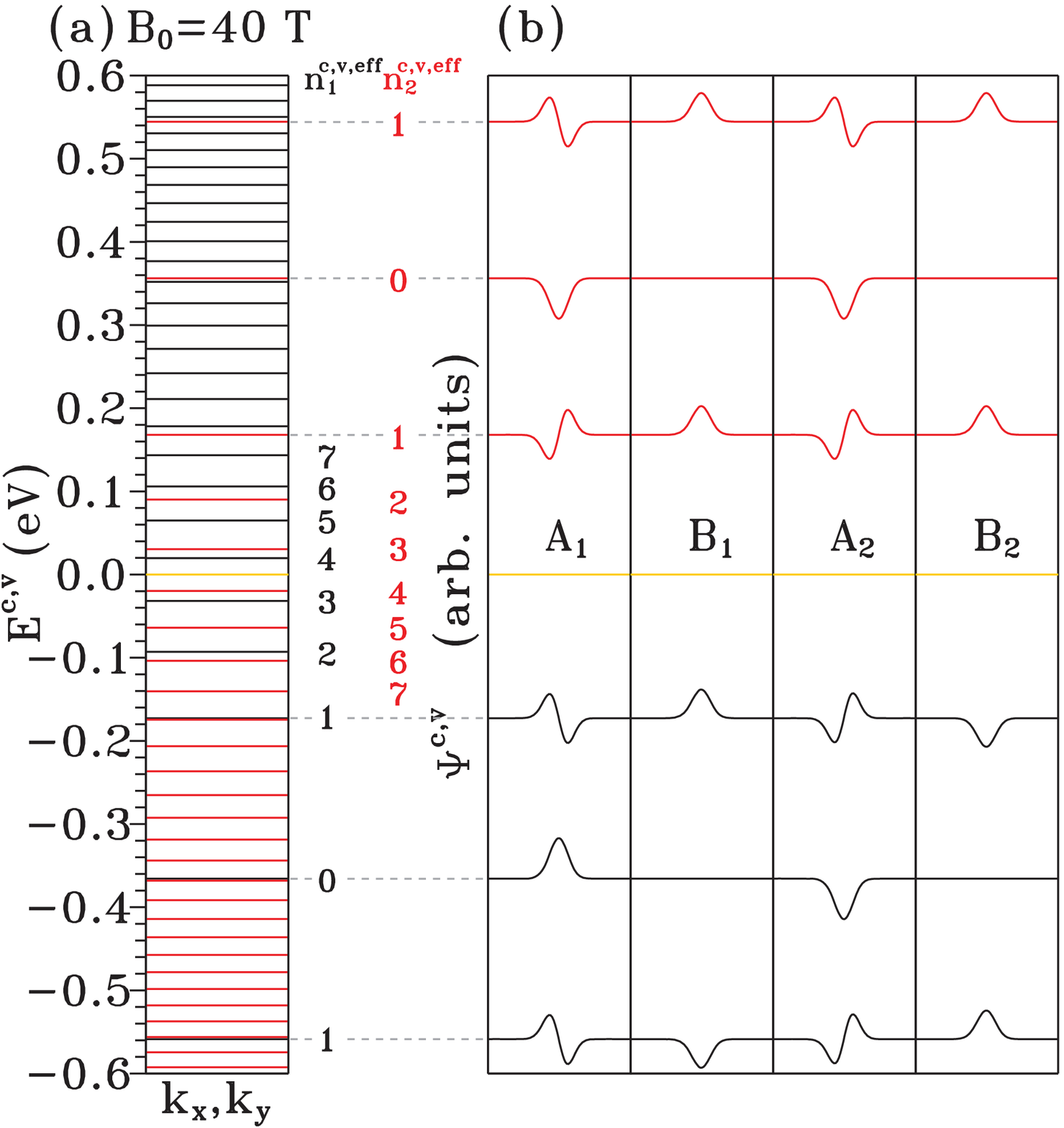}}
\end{figure}

\newpage
\begin{figure}[tbp]
\rotatebox{0}{\includegraphics[width=1\textwidth]{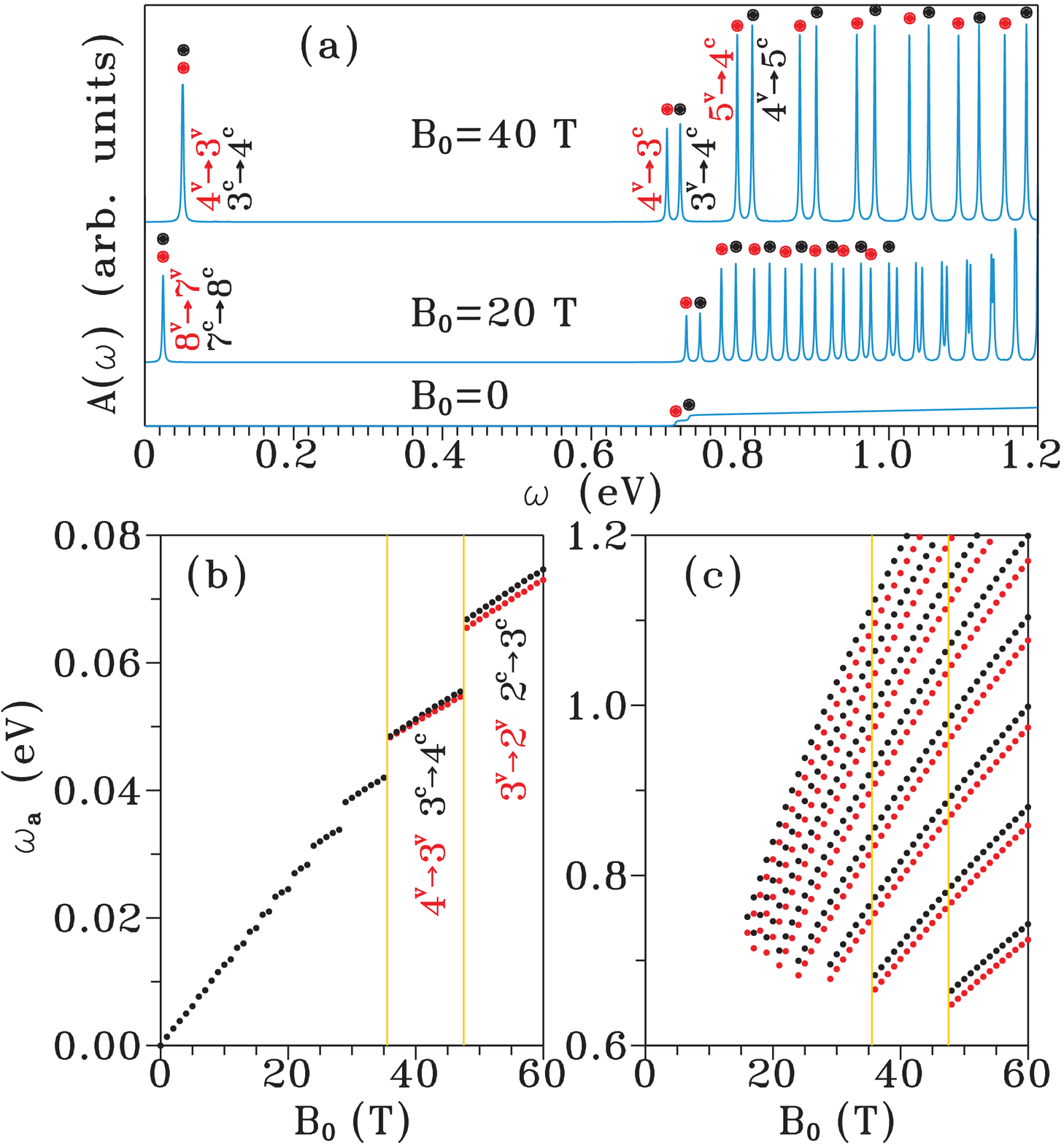}}
\end{figure}
\end{subequations}

\end{document}